\documentclass{vgtc}                          % final (conference style)
\graphicspath{{figures/}{pictures/}{images/}{./}} % where to search for the images

\usepackage{times}                     % we use Times as the main font
\usepackage{tabu}                      % only used for the table example
\usepackage{booktabs}                  % only used for the table example
\usepackage{lipsum}                    % used to generate placeholder text
\usepackage{mwe}                       % used to generate placeholder figures
\usepackage{multirow}                  % used for table formatting 
\usepackage{graphicx}                  % used for table formatting 
\usepackage{booktabs}                  % used for table formatting
\usepackage{tabularx}                  % used for table formatting

\usepackage[final]{changes}

\usepackage{mathptmx}                  % use matching math font

\usepackage{multirow}
\usepackage[table,xcdraw]{xcolor}

\newcolumntype{R}[1]{>{\raggedright\arraybackslash}p{#1}}

\onlineid{1302}

\vgtccategory{Empirical Study}

\vgtcinsertpkg

\title{Beyond Initial Creation: Two Decades of Community-based Chart Revisions from Wikimedia Commons}

\author{Jasmine Y. Shih\thanks{e-mail: jyshih@berkeley.edu}\\ %
        \scriptsize University of California, Berkeley %
\and Chase Stokes\thanks{e-mail: Chase.Stokes@morningstar.com}\\ %
     \scriptsize Morningstar, Inc. %
\and Melanie Tory\thanks{e-mail: m.tory@northeastern.edu}\\ %
     \scriptsize \centering Northeastern University %
\and Marti Hearst\thanks{e-mail: hearst@berkeley.edu}\\ %
     \scriptsize \centering University of California, Berkeley}

\teaser{
  \centering
  \includegraphics[width=\textwidth,alt={Five labeled graphics containers, each comprising consecutive versions of a chart, chart metadata (multi- or single-contributor and SVG or raster format), and edit categories labeled with LLM pipeline. Examples (a) through (d) contain two consecutive chart versions while example (e) contains four. Some of the edited chart components, such as text labels and data values, are graphically annotated and connected to their corresponding edit categories.}]{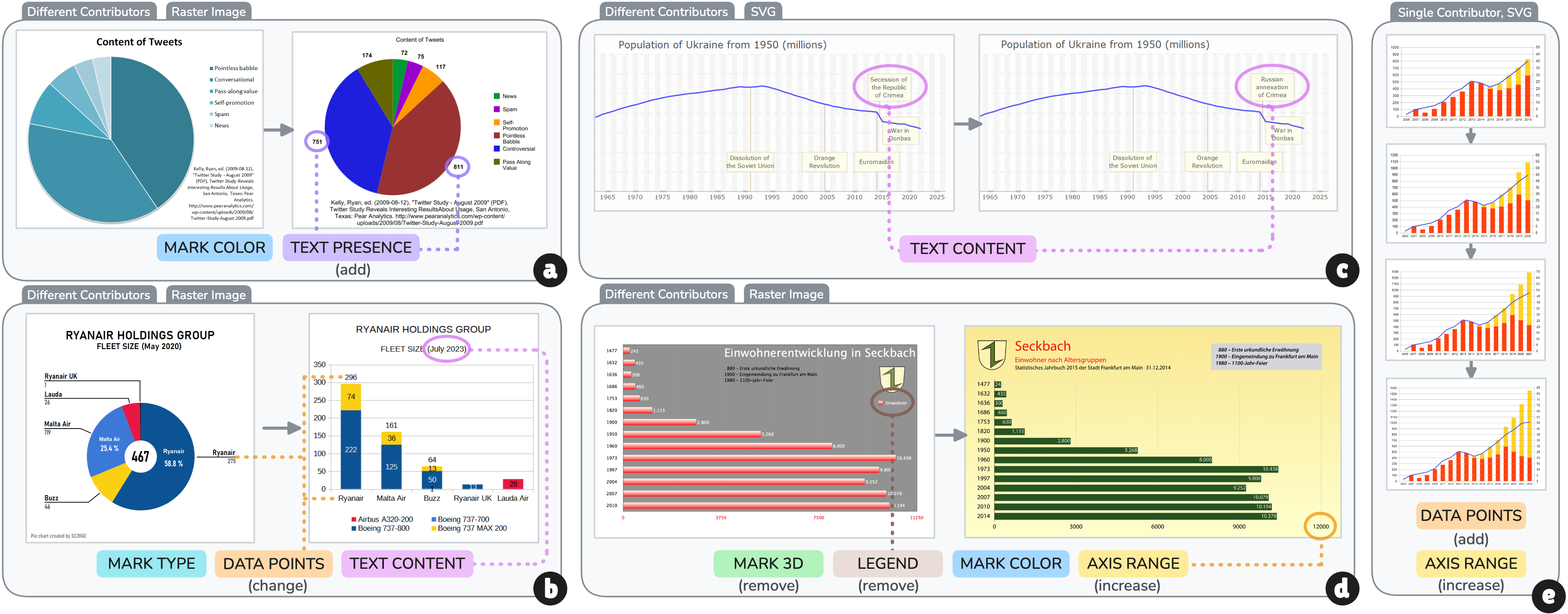}
  \caption{Examples of Wikimedia Commons chart revisions where one contributor made changes to others' work. (a) Changed the mark color scheme and added data labels. (b) Changed a donut chart to a stacked bar chart and added new data. (c) Changed the framing of a text annotation referencing a politically sensitive event. (d) Removed the 3D effects and the legend in a bar chart and increased the x-axis range. (e) A single contributor updated a chart with new data annually.}
  \vspace{-2pt}
  \label{fig:teaser}
}

\abstract{
    Open online communities like Wikipedia offer a rich space for understanding the organic evolution of public-facing artifacts, but little is known about how these communities collectively edit chart images. We investigate the phenomenon of ``community editing'' of visualizations by exploring chart revision patterns on Wikimedia Commons. We introduce a novel dataset of 7,786 longitudinal chart revisions across 1,664 images, labeled with edit categories using an LLM pipeline. Initial data analysis uncovered several novel findings. Multi-contributor revision is associated with more edits to mark styling and text content compared to single-contributor revision. Temporally, mark and text edits occur in more ``bursty'' cadences than data edits. In addition, while SVG charts attract more unique contributors and revisions than raster charts, their edit histories involve more tightly scoped and less diverse changes. This dataset enables a wide range of analyses of chart change over time.
} % end of abstract

\keywords{Community Editing, Chart Edit History, Wikimedia}

\begin{document}


%\firstsection{Introduction}

\maketitle

\section{Introduction} 

Open online communities like Wikipedia offer a rich space for understanding how data visualizations are organically maintained and iterated upon. Just as Wikipedia article text is constantly updated with new information, article chart images are revised by users in Wikimedia Commons, the media repository for Wikipedia. By analyzing public-facing chart revision activity, we can uncover how a diverse community of users collaborates---and engages in editorial conflicts---to improve visual content and design over time (see examples in \cref{fig:teaser}). Examining these in-the-wild editing behaviors provides a unique opportunity to empirically evaluate whether established visualization design principles are reflected in practice.

While the editing patterns of Wikipedia text content have been studied extensively \cite{burstinessChoi2021, spatioTemporalKiesel2017, editIntentionsYang2017}, little is known about the community-based editing of chart images. Trends in visualization authoring reflect a growing interest in tools that leverage machine learning models to facilitate chart modification \cite{plotEditGoswami2025, dynaVisVaithilingam2024}. Consequently, a recent body of work has contributed chart edit datasets primarily for model development purposes \cite{chartReformerYan2024, ChartM3Yang2025}. These existing datasets tend to be generated synthetically and therefore fail to capture the complexities in how charts evolve in unstructured, open environments. 

To address this gap, our primary contribution is a novel dataset of longitudinal community chart revisions sourced from Wikimedia Commons. Unlike the intermediate designs logged in standard analytics tools, these revisions were public-facing updates on a platform with no strict chart ownership. The dataset encompasses 7,786 revisions across 1,664 chart images. To capture the types of edits in revisions, we developed a chart edit codebook and utilized a Large Language Model (LLM) pipeline to label revision edits.

To demonstrate the richness and utility of this dataset, we conducted an initial analysis exploring three facets: contributor dynamics, temporal cadence, and file format. Our findings indicate that multi-contributor charts undergo more aesthetic edits and less structural modifications than single-contributor charts. Temporally, edits to data marks and text occur in more irregular bursts compared to data edits. Finally, we observed that charts in the SVG format are revised by more unique contributors and their revisions are more tightly scoped than charts in raster formats (e.g., .png).

This novel dataset of community-based chart revisions, along with our initial analysis results, open the door to understanding how members of an open community maintain and iterate on the content and design of data visualizations for communication purposes.\footnote{\added{The dataset is available open-access at time of publication at https://github.com/WikiVizEvolution/viz-revisions-dataset}}

\section{Related Work}

Visual media plays a crucial role in the Wikimedia ecosystem. Studies demonstrate that the images in Wikipedia articles are interacted with at higher rates than other content types \cite{largeScaleStudyRama2022}. Wikimedia Commons, the image repository for Wikipedia articles, functions as a melting pot of both community-authored and imported external media \cite{commonsWorthErickson2018}. Despite the established importance of these images, existing literature analyzing editing behaviors has largely focused on Wikipedia article text, including modeling edit intentions \cite{editIntentionsYang2017} and spatio-temporal patterns in edit reversions \cite{spatioTemporalKiesel2017}. Research studying the community curation of Wikimedia visual assets has been sparse since the platform's inception \cite{visualSideViegas2007}, and the analysis of revision dynamics has not yet been applied to data visualization images.

Although visualization literature has explored asynchronous collaboration, the majority of this work focuses on developing tools and systems to support the process of exploratory visual analytics (e.g., \cite{bringingDataKim2024, curioMoreira2025}) or discourse around visualization designs through annotations (e.g., \cite{voyagersHeer2007, manyEyesViegas2007, patinasKauer2025}). These tools prioritize data exploration or community discussion rather than the visual artifacts of shared revision work. Our research diverges from this process-centric focus by examining the visualization images created by an open community for the purpose of public communication.

Prior studies reveal that collective editing presents both benefits and costs. While collaboration iteratively improves information accuracy \cite{transcriptMunteanu2008} and visual design \cite{remoteVelloso2023}, media-intensive collaborative projects can also incur high coordination friction that result in lower-quality outcomes compared to individually authored works \cite{costHill2013}. This friction is further influenced by the technical format of the media, because the pixel-based structure of raster images makes them more difficult to revise directly than vector graphics \cite{colierBath2021}. The temporal dynamics of collective revision also varies widely, driven by peer interaction \cite{burstinessChoi2021}. Building upon these works, our study investigates how the content and design of public-facing charts may evolve based on social dynamics, chart origin, and file format.

We developed a set of edit categories informed by existing research analyzing visualization design workshops and online critiques. For instance, discussions often center on ensuring data relevance and correcting misleading representations  \cite{colVisCay2020, vizdatAlmahmoud2023}\deleted{, which necessitate tracking structural edits to data, chart type, and axes}. Design critiques often scrutinize aesthetic and visual encoding choices \cite{gulfKnoll2025, patinasKauer2025}\deleted{, informing our inclusion of data mark edits}. Community feedback also targets confusing wording, missing context, or cluttered legends to improve accessibility and clarity \cite{vizdatAlmahmoud2023, patinasKauer2025, gulfKnoll2025}\deleted{, motivating our analysis of text edits}. By operationalizing established qualitative findings, our analysis captures how common design considerations manifest as chart revisions in the wild.

\begin{figure*}[t]
  \centering
  \includegraphics[width=0.9\textwidth,alt={Three sections of diagrams laid out horizontally to demonstrate different aspects of data collection and labeling. Left: The criteria, from top to bottom, include collecting images from Wikimedia Commons categories with chart-related keywords, images with file version count greater than two, images classified as ``chart'' by custom classifier, chart images featured in at least one Wikipedia article, and charts with version count less than 25 (95th percentile). Middle: The diagram shows a simplified layout of a Wikimedia Commons image file page, which displays filename, most recent version image, file metadata, file version history, file usage in Wikimedia ecosystem, and the categories associated with the file. Right: Three-stage pipeline each showing the prompt structure, which consists of chart versions and specific instructions as input and either edit descriptions or edit categories as output. The first stage uses Gemini 3 Flash, second stage uses Gemini 3.1 Pro, and third stage uses GPT 5.2. To the right of the last stage, an additional container illustrates the LLM output validation step against human annotations.}]{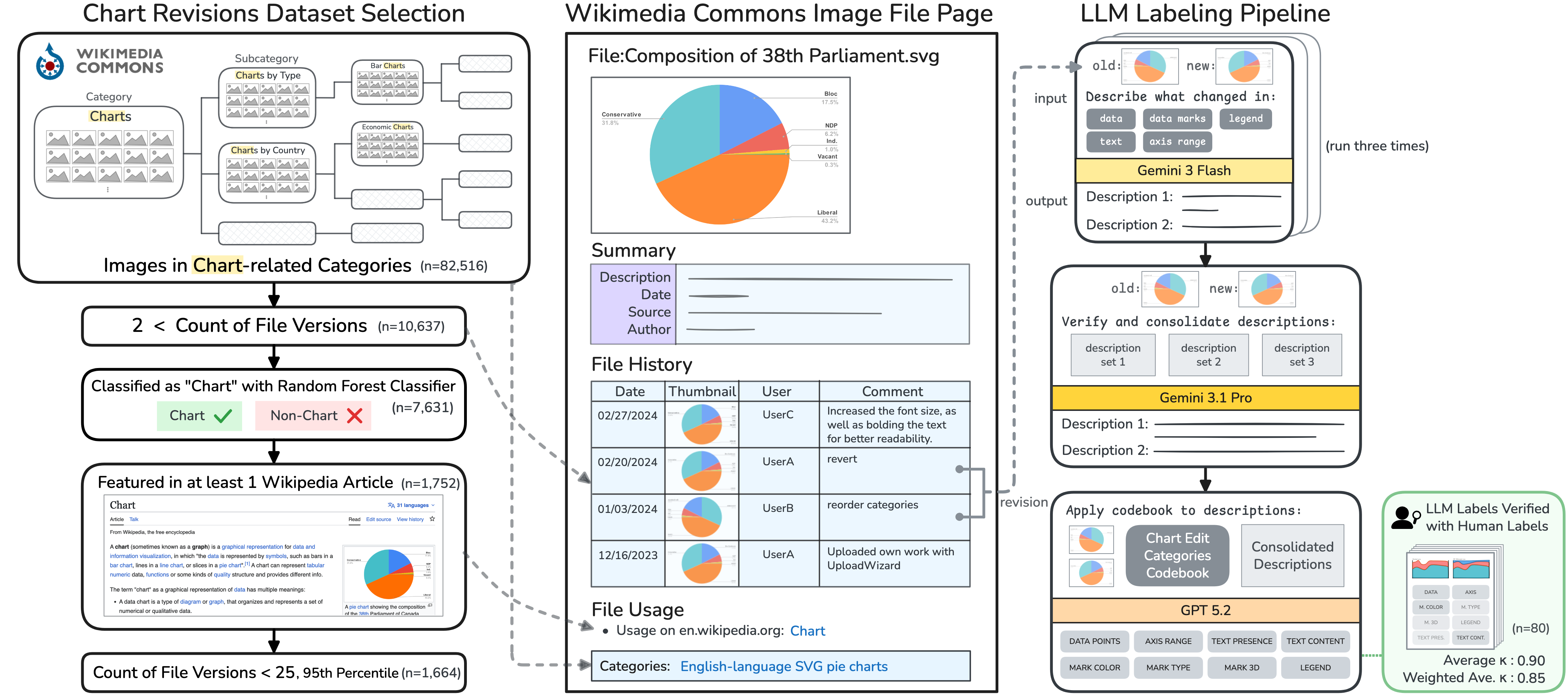}
  \caption{Data collection process. Left: Charts and their revisions were extracted from relevant Wikimedia Commons categories and filtered with several criteria. Middle: The file page of a chart image in Wikimedia Commons contains the revision history and metadata, which we collected for each chart revision. Right: Each chart revision (two consecutive versions) was labeled with edit categories using a three-stage LLM pipeline; LLM labels were verified with human annotations on a sample of 80 individual revisions, achieving $K = 0.85$.}
  \label{fig:dataset_pipeline}
  \vspace{-14pt}
\end{figure*}

\section{Dataset}
Our dataset consists of 7,786 chart revisions from the version history of 1,664 chart images created between 2004 and 2026. Each revision contains LLM-labeled edit categories and chart metadata. \deleted{\cref{fig:dataset_pipeline} illustrates the metadata collected for each revision and the data collection stages.}

\subsection{Chart Revision History Collection}

We used Wikimedia Commons' category organization to identify a pool of candidate images from over 100 million images across Wikimedia Commons. Categories can contain files and subcategories; the \textit{Charts}\footnote{https://commons.wikimedia.org/wiki/Category:Charts} category serves as both an image repository \textit{and} the root category of other related categories which may not be about statistical charts. Using the Pywikibot\footnote{https://github.com/wikimedia/pywikibot} Python library, we traversed as far as six levels down the Charts category tree, collecting images from categories whose names contain chart-related keywords (e.g., chart, graph, plot) \cite{vizByWikiLin2018}. At the time of collection (September 2025), this yielded 82,516 images from 1,659 categories.

To obtain a dataset of chart images that exhibited editing activities and were actively used in Wikipedia, we further filtered the collected images using several criteria illustrated in \cref{fig:dataset_pipeline}. We collected metadata for chart versions, including timestamp, contributor, and revision comment. SVG charts were converted to PNGs using CairoSVG\footnote{https://cairosvg.org/}\added{ for use by LLMs}. Details on the filtering methods and dataset schema can be found in supplemental material.

\subsection{Edit Category Labeling with LLMs}

The eight edit categories \added{are listed and defined} in \cref{tab:edit-categories}\deleted{comprise: \textit{Data Points}, \textit{Mark Type}, \textit{Mark Color}, \textit{Mark 3D}, \textit{Axis Range}, \textit{Text Presence}, \textit{Text Content}, and \textit{Legend}}. Each category has a set of action labels depending on the target attribute of the category (e.g., ``add'' and ``remove'' for presence\added{ of elements or effects}, ``change'' for \added{mixed edit actions or edits to }content). Each edit category, paired with an associated action, captures that a given type of change is observed in the revision.

\begin{table}[h!]
%\resizebox{\columnwidth}{!}{%
\small
\caption{Edit category definitions and proportions in revisions}
\vspace{-4pt}
\begingroup
\renewcommand{\arraystretch}{1.15} % Default value: 1
{\setlength{\tabcolsep}{3pt}
\begin{tabularx}{\columnwidth}{ll >{\raggedright\arraybackslash}X >{\raggedright\arraybackslash}p{1.75cm}} %{ll >{\raggedright\arraybackslash}X >{\raggedright\arraybackslash}X}
\hline
\textbf{Category} & \textbf{Proportion} & \textbf{Definition} & \textbf{Actions} \\ \hline
Data Points & 58.0\% & Edits to the underlying data points or series & add, remove, \deleted{both}\added{change} \\ 
Axis Range & 42.5\% & Edits to the bounds of a numeric or temporal axis & increase, decrease, \deleted{both}\added{change} \\ 
Mark Color & 15.8\% & Edits to mark fill color(s) & change \\ 
Mark Type & ~~1.8\% & Edits to chart type & change \\ 
Mark 3D & ~~0.7\% & Edits to 3D mark effects & add, remove \\ 
Text Content & 34.1\% & Edits to the words in text & change \\ 
Text Presence & 15.2\% & Addition or removal of text elements & add, remove, \deleted{both}\added{change} \\ 
Legend & ~~1.8\% & Addition or removal of legend(s) & add, remove \\ \hline
\end{tabularx}}%
\endgroup
%}
\label{tab:edit-categories}
\vspace{-5pt}
\end{table}

Building on prior chart edit taxonomies \cite{thumbnailKim2024, chartReformerYan2024}, we further grounded our edit categories in the collected chart revision dataset. We developed the codebook iteratively through open coding, axial coding, and evaluations of LLM labeling performance on intermediate category sets. Our initial edit categories included fine-grained changes to chart components and properties, including layout, dimensions, styling, and placement. \added{Due to LLM limitations, we}\deleted{We further} narrowed the set to only categories where LLMs could reliably detect changes compared to a human-coded ground-truth ($\kappa > 0.7$)\added{, leading to the omission of categories concerning positional or dimensional attributes, grid lines, and tick marks}. %More information on our chart-edit codebook can be found in \textcolor{red}{supplementary materials}.

Our large-scale labeling method drew inspiration from recent research utilizing multimodal LLMs to analyze changes across images over time \cite{visualchroniclesDeng2025}. Through iterative prompt engineering and model selection to maximize labeling quality, we developed a three-stage pipeline that generated multiple sets of edit descriptions and transformed a consolidated set of descriptions into edit category labels. \deleted{\cref{fig:dataset_pipeline} illustrates the pipeline stages.}

We evaluated this pipeline against the annotations by the first author on a sample of 80 individual revisions (half randomly sampled and half selected for edit complexity). Agreement between LLM and human labels achieved a Cohen's kappa of $0.85$ (weighted by category base rates). We then labeled the remaining revisions in our dataset with this pipeline. We also used Gemini 3 Flash to label the chart type of each image \added{($\kappa = 1.0$)}. \deleted{The model agreed with 100\% of the human-labeled chart types.} 

Details on chart-edit codebook development, LLM prompts, and evaluation results can be found in supplemental material.

\subsection{Dataset Overview}

The distribution of chart types from most to least common is Line (53\%), Bar (14\%), Area (13\%), Pie (7\%), and Other (13\%), which includes less common chart types such as scatter plots, maps, and composite charts (a mix of chart types). Compared to prior works analyzing chart type distributions on the open web \cite{beagleBattle2018}, our dataset contains higher proportions of line charts and area charts.

Charts averaged 4.7 revisions ($\sigma = 4.1$), with a right-skewed distribution (median = 3, max = 24). Revisions averaged 1.7 labeled edit categories ($\sigma = 1.3$, median = 2, max = 7). The most common category was \textit{Data Points}, whereas \textit{Legend}, \textit{Mark Type}, and \textit{Mark 3D} edits were rare. \cref{tab:edit-categories} shows  each edit type's proportion in the 7,786 revisions. 20\% of revisions did not fit in any edit categories labeled; these revisions involved edits outside the scope of our codebook after adjustments for LLM consistency, such as dimensional or positional changes to chart elements. Nonetheless, 95\% of charts have at least one revision labeled with edit categories.

The chart image metadata\added{, which enables the exploration of our research questions,} is faceted by: (1) single- vs.\ multi-contributor, (2) community-authored vs.\ externally sourced, and (3) SVG vs.\ raster formats. Although subgroups are not evenly split, each contains at least one-third of the 1,664 charts. 58\% of all charts were revised by more than one contributor, 66\% were originally authored by a community member, and 53\% are in the SVG format.

\begin{table*}[]
\def\arraystretch{1.2}
\centering
\fontsize{8}{9}\selectfont
\caption{Summary of hypotheses and results.}
\vspace{-4pt}
\begin{tabular}{R{0.88\linewidth}R{0.07\linewidth}}
\hline
\textbf{Hypothesis} & \textbf{Support} \\ \hline
\multicolumn{2}{l}{\cellcolor[HTML]{dbdbdb}\textbf{RQ1: How do edits differ between single- and multi-contributor charts?}} \\
\hline
\textbf{H1 [Unique Categories]:} Charts with more unique contributors cover a wider range of unique edit categories. & \cellcolor[HTML]{9FC5E8}Supported \\
\deleted{\textbf{H2a [Mark Color \& Text]:} Multi-contributor charts undergo more mark color and text edits than single-contributor charts.}\added{\textbf{H2a [Data \& Mark Type]:} Multi-contributor charts undergo fewer data and mark type edits than single-contributor charts.} & \cellcolor[HTML]{CFE2F3}Partial \\
\deleted{\textbf{H2b [Data \& Mark Type]:} Multi-contributor charts undergo fewer data and mark type edits than single-contributor charts.}\added{\textbf{H2b [Mark Color \& Text]:} Multi-contributor charts undergo more mark color and text edits than single-contributor charts.} & \cellcolor[HTML]{CFE2F3}Partial\\ \hline
\multicolumn{2}{l}{\cellcolor[HTML]{dbdbdb}\textbf{RQ2: How does editing cadence differ by contributor count, edit category, and chart origin?}} \\
\hline
\textbf{H3 [Contributors]:} Multi-contributor charts exhibit more bursty editing cadences than single-contributor charts. & \cellcolor[HTML]{9FC5E8}Supported \\
\textbf{H4 [Mark \& Text]:} Mark edits and text edits are performed in more bursty cadences than data edits. & \cellcolor[HTML]{9FC5E8}Supported \\
\textbf{H5 [Data \& Chart Origin]:} Data edits on externally sourced charts occur in more bursty cadences than on community-authored charts. & None \\ \hline
\multicolumn{2}{l}{\cellcolor[HTML]{dbdbdb}\textbf{RQ3: How do SVG chart revisions differ from revisions to raster charts?}} \\
\hline
\textbf{H6 [Editing Scope]:} SVG charts have more tightly scoped revisions than raster charts but cover a similar range of edit categories overall. & \cellcolor[HTML]{CFE2F3}Partial \\
\textbf{H7 [Contributors \& Revisions]:} SVG charts are revised by more unique contributors and undergo more revisions than raster charts. & \cellcolor[HTML]{9FC5E8}Supported \\
\textbf{H8 [Adoption]:} \deleted{Following Wikipedia's 2006 chart creation recommendations}\added{Since 2006}, newly created charts and revisions have increasingly favored SVG over raster formats. & \cellcolor[HTML]{9FC5E8}Supported \\ 
\hline
\end{tabular}
\label{tab:hypotheses}
\vspace{-14pt}
\end{table*}

\section{Research Questions and Results}

We explored three research questions and eight total hypotheses to examine the role of different social and technical factors in community chart editing activities. Individual hypotheses and a summary of their analysis results can be found in \cref{tab:hypotheses}.

To evaluate the hypotheses, we used generalized linear models (GLMs) with Poisson, binomial, or logistic distributions for count, proportional, and binary data, respectively. For continuous metrics, such as edit ``burstiness'' \cite{MeasuringModelingBurstyKarsai2025}, we fitted ordinary least squares (OLS) regressions, using clustered or robust standard errors to account for within-chart correlation. Across all models, we controlled for relevant covariates: chart file format, origin, age, chart type, revision count, contributor count, and power-user involvement (contributors above the 90th percentile of revision count). All $p$-values were corrected with the Holm-Bonferroni method for multiple comparisons. Regression model details are in supplemental material.

%\vspace{2pt}

\noindent \textbf{RQ1: How do edits differ between charts revised by a single contributor and charts revised by multiple contributors?}

Crowd-based production communities distribute labor across latent, specialized roles (e.g., fact updater, copy-editor) \cite{editCategoriesYang2016}. We expect Wikimedia Commons users to exhibit similar behaviors in chart editing. Each contributor to a chart may bring distinct editing tendencies, diversifying the types of edits on the chart (\textbf{H1)}.

Structural modifications such as data and mark type edits typically require contextual knowledge of the underlying data and technical proficiency with the original chart creation tool. These types of edits may therefore fall on the original chart author (\textbf{H2a)}. However, aesthetic edits to mark color and text are more accessible to additional contributors, whose individual preferences \cite{infographicHarrison2015} may increase the proportion of such edit types (\textbf{H2b)}.

\textit{\textbf{RQ1 Results}}: Charts revised by multiple contributors showed more diverse edits and higher likelihood of \textit{Mark Color} (\textit{Odds Ratio}: $OR = 1.32$) and \textit{Text Content} ($OR = 1.60$) edits than single-contributor charts ($p < .001$), although no difference was found for \textit{Text Presence} ($p = 0.16$). Multi-contributor charts were also less likely to involve \textit{Mark Type} changes ($OR = 0.45$, $p < .001$). Compared to single-contributor charts, the  likelihood of \textit{Data} edits further varied by whether the chart was originally authored by a community member (less likely, $OR = 0.54$) or sourced from an external public website (more likely, $OR = 1.48$) (both $p < .001$).

%\vspace{2pt}

\noindent \textbf{RQ2: How does editing cadence differ by contributor count, edit category, and chart origin?}

Investigating how different factors are associated with the temporal revision rhythm can reveal patterns in community coordination (e.g., regular maintenance or clustered bursts). On Wikipedia, an initial edit often triggers a reactive loop of sequential adjustments \cite{writingProcessDaxenberger2016}, suggesting that community-edited charts may exhibit more ``bursty'' cadences (\textbf{H3}).

Within edit categories, data updates tend to follow predictable real-world schedules (e.g., annual reports), while subjective design choices are more prone to consensus-building or conflict \cite{dynamicsConflictsYasseri2012}, driving quicker and more reactive mark and text revisions (\textbf{H4)}.

Chart origin may further shape data-editing cadence: community-generated artifacts often exhibit ``territoriality,'' where dedicated community members maintain a stewardship over the artifact \cite{territorialityThom2010}. On the other hand, maintenance for externally sourced charts may rely on casual contributors \cite{wikipediansPanciera2009} who happen upon outdated information (\textbf{H5)}.

\textit{\textbf{RQ2 Results}}: Charts revised by multiple contributors exhibited higher burstiness than single-contributor charts ($\beta = 0.2$, $p < .001$). Comparison of edit categories revealed that \textit{Mark} edits ($\beta = 0.09$, $p = 0.01$) and \textit{Text} edits ($\beta = 0.10$, $p < .001$) tended to occur in more irregular bursts than \textit{Data} edits. The regular schedule of data editing remained consistent across both community-authored and externally sourced charts ($p = 0.65$), suggesting that chart provenance does not significantly influence data editing cadences.

%\vspace{2pt}

\noindent \textbf{RQ3: How do SVG chart revisions differ from the revisions to charts in raster formats?}

Comparing the differences in editing activities between SVG and raster formats enables the evaluation of image format as a potential constraint in collective visualization revision. The pixel-based structure of raster images make them difficult to edit with access to both the underlying data and specific charting tools, introducing higher revision friction compared to SVGs whose DOM-structured elements can be directly edited \cite{colierBath2021}. This format difference may lead to more targeted edits in individual SVG revisions than raster revisions (\textbf{H6}). SVG editing also sets a lower barrier to entry for direct edits, potentially inviting a broader base of peripheral contributors and more frequent revisions \cite{remoteVelloso2023} (\textbf{H7)}.

The Wikipedia how-to page on chart creation has explicitly recommended the SVG format for data visualizations since 2006 \cite{WikipediaHowCreate}. This top-down community guideline, combined with the bottom-up ease of SVG editing, may drive a long-term, community-wide shift toward vector-based charting (\textbf{H8)}.

\textbf{\textit{RQ3 Results}}: SVG charts attracted more unique contributors (\textit{Incidence Rate Ratio}: $IRR = 1.15$, $p < .001$) and underwent more revisions ($IRR = 1.13$, $p < .001$) than charts in raster formats, an increase associated with the format itself rather than chart age. Despite higher editing activity, SVG revisions were more tightly scoped, exhibiting a narrower breadth of edit categories not only within individual revisions ($\beta = -0.46$, $p < .001$) but also across chart revision history ($IRR = 0.87$, $p < .001$). Charts ($\beta = 0.03$, $p = 0.02$) and revisions ($\beta = 0.04$, $p < .001$) in more recent years were slightly more likely to be in SVG than raster formats, suggesting a steady community migration toward vector graphics.

\section{Discussion}

Our results largely support prior works on community editing patterns and visualization design considerations, suggesting that Wikimedia Commons contributors exhibit similar behaviors as text-editing and visualization critique communities. In addition, several unexpected findings point to characteristics unique to community-based visualization revision. Among the edit categories, text emerged as a highly edited design element.
The distinction between text \textit{content} and text \textit{presence} edits found for \textbf{H2} implies that \textit{what} a text element should express is subject to more community refinement than \textit{whether} to include it in the chart. Given the various functions that text elements serve \cite{textFunctionsStokes2026} and the global reach of Wikipedia articles, these edits may involve neutralizing biased annotations \cite{wikigandaWessler2017} or translating labels for multilingual audiences.

Additionally, the comparison between image formats revealed that SVGs paradoxically exhibit a narrower breadth of edit categories both within individual revisions and across chart revision history. We posit this phenomenon may be a side-effect of modification methods: SVGs allow for direct DOM-based tweaks, whereas raster edits often force users to re-render the entire chart in a charting tool, potentially bundling unintended changes to default colors or text labels into a single revision. Further research is necessary to disentangle user intent from tool-driven artifacts in chart revisions.

\textbf{Limitations:} Our dataset was collected using Wikimedia Commons categories rather than extracting chart images directly from Wikipedia articles. We limited our collection to charts that appeared in articles at collection time, thus missing those that had been removed from articles. Therefore, our chart sample may not fully reflect all charts in Wikipedia articles. \deleted{In addition, we did not crawl article revision history, so our analysis did not differentiate between specific versions in a chart's revision history that actually appeared in Wikipedia and those that did not.}
The dataset may also contain a small number of non-chart images, since our image type classifier had 95\% accuracy on the test set. Similarly, a small percentage of the revision edit category labels may also be erroneous. \added{Due to LLM limitations, our current set of edit categories leaves out the aspects of a chart that affect its data-ink ratio, such as the dimensions of non-data-mark components, configuration of grid lines and tick marks, and the use of graphical annotations.}

\textbf{Future Work:} While this research presents an initial foray into the analysis of community-based chart revisions, we plan to conduct more sophisticated analyses in future work. In addition to expanding the set of edit categories, we will leverage computational methods to quantify the relative degree of revision.  A major effort will focus on analyzing full edit sequences, expanding on the pairwise analysis done thus far. This will include modeling transition probabilities and detecting reversions to prior edits and ``revision wars.''
We will also integrate chart image discussion (``Talk'') pages and co-occurrences of edit categories to uncover contributor intent. Additionally, we will examine differences between language versions to explore cultural differences in visualization revisions.

\textbf{Conclusion:} We present a novel dataset of community-based chart revisions over time collected from Wikimedia Commons and labeled with LLMs. Our initial analysis characterizes the editing activities in an open community dedicated to the revision of visualizations for public communication. We found differences in edit categories and cadences between multi- and single-contributor charts. We also identified that SVG chart revisions exhibit tighter editing scope compared to raster charts. Future research can build on these findings and this dataset to design chart editing tools for community-driven iteration on visualization content and design.

\vspace{-4pt}

\section*{Figure Credits}
%\label{sec:figure_credits}
\Cref{fig:teaser}: Images sourced from Wikimedia Commons. (a) Chart by Quillaja, Public Domain; (b) Chart by SCZMGE, CC BY 4.0; (c) Chart by Nikolai Twin, CC BY-SA 4.0; (d) Chart by ArcCan, CC BY-SA 4.0; (e) Chart by Borvan53, CC BY-SA 4.0.

\noindent \Cref{fig:dataset_pipeline}: Chart by Morpheus7377, Public Domain, Wikimedia Commons.

\bibliographystyle{abbrv-doi}

\bibliography{template}

% TBD.}

\end{document}